\documentclass[aps,prd,showpacs,onecolumn,nofootinbib,superscriptaddress,amsmath,amssymb]{revtex4-2}

\usepackage{subcaption}

\usepackage[normalem]{ulem}

\usepackage{soul}
\usepackage{booktabs}
\usepackage{multirow}
\usepackage{microtype}
\usepackage{graphicx}
\usepackage{graphics}
\usepackage{color}
\usepackage{epsfig}
\usepackage{definitions}%
\usepackage{dsfont,amsmath,bm,braket}
\newcommand{\rr}[1]{{#1}}

\def\Red#1#2{{#1}}
\newcommand{\rz}[1]{\textcolor{red}{}}

\begin{document}
\title{Hall conductivity as topological invariant
	in phase space}

\author{I.V. Fialkovsky}
\email{ifialk@gmail.com}
\affiliation{Physics Department, Ariel University, Ariel 40700, Israel}
\affiliation{CMCC-Universidade  Federal  do  ABC,  Santo  Andre,  S.P.,  Brazil}

\author{M. Suleymanov}
\email{michaels@ariel.ac.il}
\affiliation{Physics Department, Ariel University, Ariel 40700, Israel}

\author{Xi Wu}
\email{wuxi@ariel.ac.il}
\affiliation{Physics Department, Ariel University, Ariel 40700, Israel}

\author{C.X. Zhang}
\email{zhang12345s@sina.com}
\affiliation{Physics Department, Ariel University, Ariel 40700, Israel}

\author{M.A. Zubkov \footnote{On leave of absence from Institute for Theoretical and Experimental Physics, B. Cheremushkinskaya 25, Moscow, 117259, Russia}}
\email{zubkov@itep.ru}
\affiliation{Physics Department, Ariel University, Ariel 40700, Israel}

\date{\today}

\begin{abstract}
It is well known that the quantum Hall conductivity in the presence of constant magnetic field is expressed through the topological TKNN invariant. The same invariant is responsible for the intrinsic anomalous quantum Hall effect (AQHE), which, in addition, may be represented as one in momentum space composed of the two point Green's  functions. We propose the generalization of this expression to the QHE in the presence of non-uniform magnetic field. The proposed expression is the topological invariant in phase space composed of the Weyl symbols of the two-point Green's  function. It is applicable to a wide range of non-uniform tight-binding models, including the interacting ones.
\end{abstract}


\maketitle

\section{Introduction}

The topological nature of Hall conductivity is typically associated with the TKNN invariant \cite{Thouless1982}, which has been proposed for the systems subject to   constant external magnetic field. This invariant is also relevant for the description of intrinsic anomalous quantum Hall effect in homogeneous systems. In \cite{Thouless1982} the Hall conductivity has been expressed
as an integral of Berry curvature $\bA$ over the magnetic Brillouin zone
\be
\si_{xy}=\frac{e^2}{h}\frac1{2\pi} \int d^2k [{\bm\nabla}\times \bA(\bk)]
\ee
$$
\bA(\bk) = -\ii \braket{u(\bk)|{\bm \nabla}|u(\bk)}.
$$
where $\bk = (k_1,k_2)$, while $[{\bm\nabla}\times \bA(\bk)] = \epsilon^{ij}\nabla_i A_j$.
{Speaking mathematically}, this expression represents the first Chern class of the $U(1)$ principal fiber bundle on  the Brillouin zone. Its topological nature can be recognized by the following naive argument. Since the compact Brillouin zone does not have boundary, the application of Stokes theorem would give zero if $\bA$ is uniquely defined on the entire Brillouin zone. The nontrivial topology makes only integer values possible (see, e.g. \cite{Kaufmann:2015lga} and references therein). For the discussion of topology related to the TKNN invariant see also \cite{Avron1983,Fradkin1991,Tong:2016kpv,Hatsugai1997,Qi2008}.

The disadvantage of the TKNN invariant is that its application is limited to the systems with constant magnetic field or homogeneous quantum Hall insulators. Moreover, in the presence of interactions this invariant is not defined at all. The latter problem may be solved using the alternative form of the TKNN invariant, in which it is expressed through the two point Green's  function. The latter is well defined in the presence of interactions, which allows to define the corresponding topological invariant for the systems with interactions.

The simplest topological invariant composed of the two-point Green's  function is responsible for the stability of the Fermi surface in the $3+1$ D systems:
\be
N_1= \tr \oint_C \frac{1}{2\pi \ii}
G(p_0,\bp)d G^{-1}(p_0,\bp).
\ee
Here $C$ is an arbitrary contour, which encloses the Fermi surface \cite{Volovik2003a} in four-dimensional momentum space.
Similarly, the topological stability of Fermi points is protected by \cite{Matsuyama1987a,Volovik2003a}
\be
N_3=\frac1{24\pi^2} \epsilon_{\mu\nu\rho\lambda} \tr\int_S dS^\mu
G\partial^\nu G^{-1}
G\partial^\rho G^{-1}
G\partial^\lambda G^{-1}.
\ee
Here $S$ is the surface encompassing all the Fermi points.
These invariants were shown to be applicable for the interacting systems, but the  non-homogeneous ones are still out of their scope.

It has been demonstrated that in the absence of the inter-electron interactions the TKNN invariant for the intrinsic QHE (existing without external magnetic field) may be expressed through the momentum space Green's function \cite{IshikawaMatsuyama1986,Volovik1988} (see also Chapter 21.2.1 in \cite{Volovik2003a}). In \cite{Zubkov2018a} the alternative derivation of this expression has been proposed using the Wigner - Weyl formalism, it was later repeated independently and extended to multi - dimensional space - time in  \cite{mera2017topological}.
For the $2+1$D fermions (embedded in $3+1$D space-time) \Red{}{in the presence of electric field along} the Hall conductivity is given by
$$
\sigma_H = \frac{\cal N}{2\pi},
$$
where
{\be
{\cal N}
=  \frac{ \epsilon_{ijk}}{  \,3!\,4\pi^2}\, \int d^3p \Tr
\[
{G}(p ) \frac{\partial {G}^{-1}(p )}{\partial p_i}  \frac{\partial  {G}(p )}{\partial p_j}  \frac{\partial  {G}^{-1}(p )}{\partial p_k}
\].
\label{calM2d230}
\ee}%
This expression is topological invariant in momentum space, i.e. it is not changed when the given system is modified smoothly. Originally this representation for $\sigma_H$ was  derived for the non-interacting systems. It is widely believed, however, that in the presence of interactions the expression of    \cite{IshikawaMatsuyama1986,Volovik1988} remains valid, if the non-interacting two-point Green's  function has been substituted by full  two-point Green's function with the interaction corrections. In the $2+1$D QED this has been proved in \cite{parity_anomaly,parity_anomaly_}. The corresponding property is now referred to as non-renormalization of the parity anomaly in $2+1$D Quantum Electrodynamics by the higher orders of perturbation theory.
This is actually the proof that \Red{}{the proof that} the anomalous quantum Hall (AQHE) conductivity in relativistic $2+1$ QED does not have radiative corrections.
In a recent paper \cite{ZZ2019} the influence of interactions on the anomalous quantum Hall (AQHE) conductivity in the tight-binding models of $2 + 1$D topological insulator and $3 + 1$D Weyl semimetal has been investigated. Several types of interactions were  considered including the contact four-fermion interactions, Yukawa and Coulomb ones . It was shown that in the one-loop approximation (i.e. in the leading order) the Hall conductivity for the insulator is the topological invariant, which is given by the expression {of Eq. (\ref{calM2d230})}  \cite{IshikawaMatsuyama1986,Volovik1988} composed of the complete two-point Green's  function of the interacting model.
However, this work only studied anomalous quantum Hall conductivity in the absence of magnetic field. Moreover, higher-order effects of the interaction were not taken into account. It is worth mentioning, that the influence of interactions on the Hall conductivity in external magnetic field has been discussed widely in the past (see, for example \cite{KuboHasegawa1959,Niu1985a,Altshuler0,Altshuler} and references therein), however, those considerations have been limited by the case of constant magnetic field.

In the present paper we review the results obtained with the participation of the authors on the
Hall conductivity in the non-homogeneous systems including those with varying magnetic field. The main result of this study as we see it is the proposition \cite{ZW2019} of a new expression for the Hall conductivity. It is a topological invariant composed of the Wigner transformed two-point Green's  functions.  This proposition has been developed in \cite{FZ2019}, where the condensed matter systems with $\mathds{Z}_2$ invariance (graphene, in particular) were considered in the presence of elastic deformations. Besides, in \cite{ZZ2019_2} the proof was presented that in the presence of interactions the Hall conductivity is still given by the expression proposed in \cite{ZW2019}, in which the interacting two point Green's  function is substituted. The whole development reviewed in the present paper is based on the version of the lattice Wigner - Weyl calculus summarized in \cite{SZ2018}.

The Wigner-Weyl formalism was proposed originally by H. Groenewold \cite{Groenewold1946} and J. Moyal \cite{Moyal1949} in the context of the one-particle quantum mechanics. The main notions of this formalism are the Weyl symbol of operator and the Wigner transformation of function. This calculus accumulated the ideas of H. Weyl \cite{Weyl1927} and E. Wigner \cite{Wigner1932}. In quantum mechanics the Wigner-Weyl formalism utilises instead of wave function the so called Wigner distribution, which is the function of both coordinates and momenta. The operators of physical quantities are described by their Weyl symbols. The product of operators in this calculus becomes the Moyal product of their Weyl symbols \cite{Ali2005,Berezin1972}.
Wigner-Weyl calculus has been widely applied in quantum mechanics \cite{Curtright2012,Zachos2005}.

In the recent decades various modifications of Wigner-Weyl formalism were proposed \cite{Cohen1966,Agarwal1970,E.C.1963,Glauber1963,Husimi1940,Cahill1969,Buot2009}.
In particular, the Wigner-Weyl formalism was modified in order to be applied to the Quantum Field Theory. The analogue of the Wigner distribution was introduced in QCD \cite{Lorce2011,Elze1986}. It has been used in the field-theoretic kinetic theory \cite{Hebenstreit2010,Calzetta1988}, in noncommutative field theories \cite{Bastos2008,Dayi2002}. Certain applications of the Wigner-Weyl formalism were proposed to several fields of theoretical physics including cosmology \cite{Habib1990,Chapman1994,Berry1977}.

In the works of one of the authors of the present paper the Wigner-Weyl formalism has been applied to the study of the nondissipative transport phenomena \cite{Zubkov2017,Chernodub2017,Khaidukov2017,Zubkov2018a,Zubkov2016a,Zubkov2016b}. In particular, it was shown that the response of nondissipative currents to the external field strength is expressed through the topological invariants that are robust to the smooth deformations of the system.
The absence of the equilibrium chiral magnetic effect \cite{Kharzeev2014} was demonstrated within the lattice regularized field theory \cite{Zubkov2016a}. The anomalous quantum Hall effect was studied in \cite{Zubkov2016b}. The chiral separation effect was derived \cite{Metlitski2005} within the lattice models \cite{Khaidukov2017,Zubkov2017}. The same method was also applied to the investigation of the phases of high density QCD \cite{Zubkov2018a}. In addition, the scale magnetic effect \cite{Chernodub2016} was considered using the same technique \cite{Chernodub2017}.

It is worth mentioning, that momentum space topological invariants were widely used in the context of condensed matter physics theory
\cite{HasanKane2010,Xiao-LiangQi2011,Volovik2011,Volovik2007,VolovikSemimetal}. They appear to be responsible for the topological protection of gapless fermions at the edges of the topological insulators \cite{Gurarie2011a,EssinGurarie2011} and the gapless fermions in Weyl semi-metals \cite{Volovik2003a,VolovikSemimetal}. The fermion zero modes related to topological defects in $^3$He are described by momentum space topology as well \cite{Volovik2016}. In the context of elementary particle theory the topological invariants in momentum space were considered in \cite{NielsenNinomiya1981a,NielsenNinomiya1981b,So1985a,IshikawaMatsuyama1986,Kaplan1992a,Golterman1993,Volovik2003a,Hovrava2005,Creutz2008a,Kaplan2011}.

The paper is organized as follows. In Sect. \ref{SectWigner} we recall the basic notions of Wigner-Weyl calculus. In Sect. \ref{Sec:GroEq} we express electric current through the Wigner transformation of the two-point Green's  function. In Sect. \ref{SectCond} we present the expression for Hall conductivity through the topological invariant in phase space composed of the Wigner transformed Green's  function. In Sect. \ref{SectKubo} we rederive the expression for the Hall conductivity using Kubo formula. In Sect. \ref{SectInt} the role of inter-fermion interactions is discussed. In Sect. \ref{SectElast} we discuss the systems with elastic deformations. In Sect. \ref{SectGroenewold}  we give the iterative solution of the Groenewold equation. In Sect. \ref{SectConcl} we end with the conclusions.

\section{Wigner-Weyl formalism}

\label{SectWigner}

We start with a very brief introduction to the Wigner-Weyl formalism. We put aside many questions, including alternative formulations and recent developments of the formalism such as those that gave rise to the rapidly developing area of deformation quantization.

Wigner-Weyl formalism was proposed almost as early as the operator formulation of Quantum Mechanics (QM), but without operators and Hilbert spaces \cite{Weyl1927,Wigner1932,Groenewold1946,Moyal1949}. It can be understood as the correspondence between the QM operators and the functions in phase space,
$$
\hat A\equiv A(\hat \bx, \hat \bp)
	\quad\leftrightarrow \quad
	A_W\equiv A_W(\bx,\bp),
$$
such that
\be
	(\hat A \hat B)_W = A_W \ast B_W,
\label{A*B}
\ee
\be
	\tr \hat A = \Tr A_W
\label{tr=Tr}
\ee
\be
	\Tr(A_W\ast B_W) = \Tr(A_W B_W)
\label{no-star}
\ee
with appropriate definitions for $\ast$--product (associative, non-commutative) and $\Tr$ operation.

Actually, the formalism can be formulated totally independent from ordinary QM, with Schrodinger equation replaced by the Moyal equation
$$
\frac{\partial \rho}{\partial t} = \frac{H\ast \rho-\rho\ast H }{\ii \hbar}\equiv
\{\!\!\{ H,\rho\}\!\!\}.
$$
Here $\rho$ is the Wigner function, i.e. Weyl symbol of the density matrix $\hat \rho$.
For an arbitrary operator $\hat A$ in {D dimensional} continuous theory the Weyl symbol can be defined as
\be
A_W(\bx,\bp)
= \frac1{(2\pi\hbar)^D}\int d^D q \,e^{\ii \bq \bx/\hbar}\braket{\bp+\bq/2|\hat A|\bp-\bq/2}.
\label{A_W}
\ee
Notice, that the definition of a similar object for the lattice models is accompanied by certain difficulties (see discussion below). Moyal product is defined as
\be
\ast = e^{\tfrac{\ii\hbar}2(\overleftarrow{\partial}_\bx\overrightarrow{\partial}_\bp-\overleftarrow{\partial}_\bp\overrightarrow{\partial}_\bx)}.
\label{star}
\ee
The trace operation $\Tr$ then stands for the integration over the whole phase space and summation over the inner indices, if any
\be
\Tr A_W(\bx,\bp )\equiv
	\int d^Dx d^Dp \tr A_W(\bx,\bp ).
	\label{Tr}
\ee
The above property of \Ref{tr=Tr} may be easily proved in this case.

On the lattice we define Weyl symbol of operator $\hat A$ as:
\be
A_W(\bx,\bp )
	=
	\int_{{\cM}}d^D\cP e^{i\bx\bcP} \bra{\bp+\tfrac\bcP{2}} \hat{A}  \Ket{\bp-\tfrac\bcP{2}}.
\ee
The $\bcP$-integral goes over the first Brillouin zone ${\cM}$ corresponding to the lattice in configuration space, i.e. in ${\cM}$ we identify the points that differ by vectors of reciprocal lattice, $\bgj$.

Now let us consider the Weyl symbol $(AB)_W(\bx,\bp )$ of the product of two operators $\hat A$ and $\hat B$ such that their matrix elements $\Bra{\bp+\frac\bq {2}} \hat{A} \Ket{\bp-\frac\bq {2}}$ and $\Bra{\bp+\frac\bq {2}} \hat{B} \Ket{\bp-\frac\bq {2}}$ are nonzero only when $\bq$ remains in the small vicinity of zero. Then
\begin{equation}\begin{aligned}
&(AB)_W(\bx,\bp )=
\int_{{\cM}} d^D\cP \int_{\cM} d^D\cR
e^{\ii \bx \bcP}
\Bra{\bp+\tfrac\bcP{2}} \hat{A} \Ket{\bcR}
\Bra{\bcR} \hat{B} \Ket{\bp-\tfrac\bcP{2}}\\
&=\frac{1}{2^D}\int_{{\cM}} d^D\cP d^D\cK
e^{\ii \bx \bcP}
\Bra{\bp+\tfrac\bcP{2}} \hat{A} \Ket{\bp-\tfrac\bcK{2}}
\Bra{\bp-\tfrac\bcK{2}}\hat{B} \Ket{\bp-\tfrac\bcP{2}}\\
&= \frac{2^D}{2^D }\int_{{\cM}} d^Dq d^Dk e^{\ii \bx (\bq+ \bk)}
\Bra{\bp+\tfrac\bq {2}+\tfrac{ \bk}{2}} \hat{A} \Ket{\bp-\tfrac\bq {2}+\tfrac{ \bk}{2}}
\Bra{\bp-\tfrac\bq {2}+\tfrac{ \bk}{2}}\hat{B} \Ket{\bp-\tfrac\bq {2}-\tfrac{\bk}{2}}\\
&= \int_{{\cM}} d^Dq d^Dk
\[  e^{\ii \bx  \bq}
\Bra{\bp+\tfrac\bq {2}} \hat{A} \Ket{\bp-\tfrac\bq {2}}
\]
e^{\tfrac{ \bk}{2}\cev{\partial}_\bp-\tfrac\bq {2}\vec{\partial}_\bp}
\[  e^{\ii \bx   \bk}
\Bra{p+\tfrac{ \bk}{2}}\hat{B} \Ket{\bp-\tfrac{ \bk}{2}}
\]\\
&= \[ \int_{{\cM}} d^Dq  e^{\ii \bx  \bq}
\Bra{\bp+\tfrac\bq {2}} \hat{A} \Ket{\bp-\tfrac\bq {2}}
\]
e^{\tfrac{\ii}{2} \(- \cev{\partial}_\bp\vec{\partial}_\bx+\cev{\partial}_{\bx}\vec{\partial}_\bp\)}
\[ \int_{{\cM}} d^Dk e^{\ii \bx   \bk}
\Bra{\bp+\tfrac{ \bk}{2}}\hat{B} \Ket{\bp-\tfrac{ \bk}{2}}
\].
\label{Z}
\end{aligned}
\end{equation}
Here the bra- and ket- vectors in momentum space are defined modulo vectors of reciprocal lattice $\bgj$,  as it is inflicted by the periodicity of the lattice. In the second line we changed variables
$$
\bcP = \bq+ \bk , \quad \bcK = \bq- \bk
$$
$$
\bq = \frac{\bcP+\bcK}{2}, \quad  \bk =\frac{\bcP-\bcK}2
$$
with the Jacobian
$$
J = \left|\begin{array}{cc} 1 & 1 \\
-1 & 1 \end{array} \right| = 2^D.
$$
This results in the factor ${2^{D}}$ in the third line. Here $D$ is the dimension of space. In the present paper it may be either $2$ or $3$.

Hence, the  Moyal product  may be defined similar to the case of continuous space
\begin{equation}\begin{aligned}
&(AB)_W(\bx,\bp )=
A_W(\bx,\bp )
e^{\frac{\ii}{2} \( \cev{\partial}_{\bx}\vec{\partial}_\bp-\cev{\partial}_\bp\vec{\partial}_{\bx}\)}
B_W(\bx,\bp ).
\label{ZAB}\end{aligned}\end{equation}
Notice, that for the chosen form of Wigner transformation on a lattice the above equality is approximate and works only if the operators $\hat{A}$, $\hat{B}$ are close to diagonal.

In practice, the above lattice Weyl symbol works as a good approximation for the systems subject to slowly varying external electromagnetic field and/or in the presence of weak elastic deformations \cite{SZ2018,FZ2019}. In particular, the value of external magnetic field $B$  should be much smaller than $1/a^2$ (where $a$ is the typical lattice spacing), i.e. $B \ll {10000}$ Tesla for the real crystal lattices. In the artificial lattices, this value may be much smaller, see e.g. \cite{Wang}.

An important consequence of the formalism is the Groenewold equation relating the Weyl symbols of the lattice Dirac operator, $\hat Q$, and its Green's function, $\hat G$. At the operator level they are simply inverse,
\be
	\hat Q \hat G = {\mathds 1}.
\ee
Calculating Weyl symbol of both sides and using \Ref{A*B}, we obtain
\be
	(\hat Q \hat G)_W
	= Q_W\ast G_W = 1.
	\label{GreoEqu}
\ee
This equation can be solved iteratively, for the detailed treatment see \cite{SZ2018}. For the purposes of the present paper we will only need the obvious first approximation in the derivative expansion
\be
	G_W\approx G_W^{(0)} - G_W^{(0)}*Q_W^{(1)}*G_W^{(0)},
\ee
valid for $ Q_W\approx Q_W^{(0)}+ Q_W^{(1)}$. Here $Q_W^{(1)}$ is linear in the external electric field.

\section{Wigner-Weyl field theory}

\label{Sec:GroEq}

In this section we closely follow \cite{ZW2019,FZ2019,ZZ2019_2}. Partition function of a general model with fermions may be written in Euclidian space - time as
\be
Z = \int D\bar{\Psi}D\Psi
\,\, e^{S[\Psi,\bar{\Psi}]}
\label{Z01}
\ee
with the action
\be
	S[\Psi,\bar{\Psi}]= \int d^{D+1}p \, d^{D+1}q \,
\bar{\Psi}^T(p)\,{Q}(p,q)\,\Psi(q),
\ee
where the integration measure and normalization are understood to be chosen appropriately for model under consideration.

As usual, the Dirac operator, $\hat{Q}$, and its inverse, the Green's function, $\hat{G} = \hat{Q}^{-1}$, acting in the Hilbert space ${\cal H}$ are related by
\be
\hat{Q} \hat{G} = 1
\label{QG=1}
\ee
or, equivalently,
$$
\langle p|\hat{Q}\hat{G}|q\rangle = \delta^{(D+1)}({p}-q ).
$$
in terms of their matrix elements. $D$ is the dimensionality of space. The basis of $\cal H$ is normalized as
$\langle p| q\rangle = \delta(p_{D+1}-q_{D+1})\delta^{(D)}(\bp-\bq)$.
One can see that the action may be represented as the trace of the product of operators
\be
S[\Psi,\bar{\Psi}] = \tr \(\hat{W}[\Psi,\bar{\Psi}] \hat Q\),
\ee
where $\hat W[\Psi,\bar{\Psi}]$ is the Wigner operator
\be
\hat W=\ket\Psi\bra\Psi.
\ee
Variation of partition function may then be written as
\be
\delta Z = \int D\bar{\Psi}D\Psi \, e^{S }
	\, \tr\(\hat{W} \delta \hat Q\)
	= Z \tr\(\braket{\hat W} \delta \hat Q\),
\ee
where the usual vacuum expectation value was used,
\be
\langle \hat O \rangle = \frac1Z\int D\bar{\Psi}D \Psi \, \hat O e^{S[\Psi,\bar{\Psi}]}.
\ee
Further employing Peierls substitution, i.e. noting that introduction of an electromagnetic (EM) potential $A$ simply shifts the momenta, $p\to p-A(x)$,  we obtain for the slowly varying $A$:
\be
	\delta \hat Q = - \partial_{p_k}\!\hat Q\, \delta A_k,
\ee
and using the basic Weyl transformation properties \Ref{tr=Tr} and \Ref{no-star}, we come to
\be
\begin{split}
\delta Z &	=  Z\int \frac{d^{D+1}p}{(2\pi)^{D+1}} d^{D+1}x \,\tr\( G_W(x,p) * \partial_{p_k}Q_W(x,p) \delta A(x)\)\\
	&
	=  Z \int d^{D+1}x \delta A(x) \int \frac{d^{D+1}p}{(2\pi)^{D+1}}\, G_W(x,p) \partial_{p_k}Q_W(x,p).
	\nonumber
\end{split}
\ee
Thus the current density is
\be
	\braket{J_k(x)} = -\int \frac{d^{D+1}p}{(2\pi)^{D+1}} \, G_W(x,p) \partial_{p_k}Q_W(x,p).
\label{j(x)}
\ee
Note that it is not the topological invariant, it must be averaged over the whole area/volume of the sample to have this property. Indeed, the total current,
\be
	\bar{J_k} \equiv \int d^{D+1}x \braket{J_k(x)} = -\Tr( G_W \ast \partial_{p_k} Q_W),
\ee
is topological invariant: under the small variations of lattice Dirac operator the Weyl symbol, $ Q_W\approx Q_W + \delta Q_W$, the Green's function varies accordingly, $G_W\approx G_W +   \delta G_{W}$, and then
$$
\delta\( \Tr\[G_W \ast \partial_{p_k} Q_W \] \)
=  \Tr\[G_W \ast  \partial_{p_k} \delta Q_W +\delta G_{W} \ast\partial_{p_k} Q_W  \].
$$
Given that $ \delta G_{W}  = -G _W\ast \delta Q_W \ast G_W  $, which follows from Eq. \Ref{GreoEqu}, the latter two terms become
\be
\begin{split}
	\Tr&\[ G_W \ast  \partial_{p_k} \delta Q_W-G _W*\delta Q_W*G
	\ast\partial_{p_k}  Q_W  \] \\
	& = \Tr\[ G _W\ast \partial_{p_k} Q_W  \ast G _W\ast \delta Q_W-G _W
	\ast \delta Q_W\ast G _W\ast\partial_{p_k} Q_W  \] 	\nonumber\\	
\end{split}
\ee
where we integrated by parts and used that $ \partial_{p_l}G_{W}  = -G \ast \partial_{p_l} Q_W  \ast G  $. Now the simple cyclic transformation inside the trace proves that
\be
\delta \bar{J_k} =0.
\label{deltaJ}
\ee

\section{Conductivity}

\label{SectCond}
To obtain expression for the conductivity let us consider the current density \Ref{j(x)}, and represent the electromagnetic field $A$ as a sum of the two contributions:
$$
	A = A^{(M)} + A^{(E)}
$$
where $A^{(E)}$ is responsible for the constant external electric field while $A^{(M)}$ produces the magnetic field and contains the electric potential of impurities. Provided that the former contribution is  weak, we obtain
\be
\braket{J(x)}
\equiv - \int \frac{d^{D+1}p}{(2\pi)^{D+1}} \, G_W(x,p) \partial_{p_k}Q_W(x,p)
	\approx  j^{(0)}+j^{(1)}_l   A^{(E)l}+
 j^{(2)}_{lm}
F^{(E)lm}+\ldots
	\label{Jappr}
\ee
The first term here is expected to be zero if the Bloch theorem is valid (which occurs for the majority of the systems discussed here), while the second should be absent due to the gauge invariance.

To calculate $j^{(2)}_{lm} $, which eventually defines the conductivity, we recall, that for the slowly varying field $A$ the expression for $Q_W$ may be represented as \cite{SZ2018}
\be
	Q_W \approx  Q_W^{(0)}-  \partial_{p_m} Q_W^{(0)} A^{(E)}_m .
	\label{Qappr}
\ee
The Groenewold equation \Ref{GreoEqu} that relates $G_W$ and $Q_W$ has the form
$$
	G_W * Q_W = 1.
$$
It can be solved iteratively giving
\be
	G_W\approx G_W^{(0)} + G_W^{(0)}\ast(\partial_{p_m} Q_W^{(0)} A^{(E)}_m) \ast G_W^{(0)}.
	\label{Gappr}
\ee
Further expanding the stars in the above expression, which contains the derivatives in $x$ acting on $A^{(E)}$, we have
$$
G_W\approx G_W^{(0)}+  G_{W,m}^{(1)} A^{(E)}_m + 	G_{W,lm}^{(2)}\rr{ \partial_{[l} A^{(E)}_{m]}},	
$$
where
$$
	G_{W,m}^{(1)} = G_W^{(0)}\ast \partial_{p_m} Q_W^{(0)} \ast G_W^{(0)},\qquad
	G_{W,lm}^{(2)} = \frac\ii2 G_W^{(0)}\ast \partial_{p_l} Q_W^{(0)} \ast   G_W^{(0)}
	\ast \partial_{p_m} Q_W^{(0)} \ast   G_W^{(0)}.
$$
Upon substitution of \Ref{Qappr} and \Ref{Gappr} into \Ref{Jappr} we obtain
\be
		\braket{J_k(x)}
			\approx \rr{-}\frac{\ii  F^{(E)}_{lm }(x)}2 \int \frac{d^{D+1}p}{(2\pi)^{D+1}}
			\,\tr \(
			G_W^{(0)} \ast \partial_{p_l} Q_W^{(0)} \ast
			G_W^{(0)} \ast \partial_{p_m} Q_W^{(0)} \ast
			G_W^{(0)} \boldsymbol{\cdot} \partial_{p_k} Q_W^{(0)}
			\),
\label{J(x)}			
\ee
where the last product is an ordinary one. Notice that $Q_W^{(0)}=Q_W^{(0)}(x,p)$,  $G_W^{(0)}=G_W^{(0)}(x,p)$. Assuming that the external field is constant across the system, $F^{(E)}_{lm}=const$, one may calculate the total current averaged over the area $\cA$ and also averaged in time
\be
	{\mathcal J}_k \equiv \frac1{\beta\cA}\int d^{D+1}x \braket{J_k(x)}
	= {\mathcal W}_{lmk}F^{(E)}_{lm } ,
	\label{cJ}
\ee
\be
	{\mathcal W}_{lmk}
	\equiv \rr{-}\frac{\ii  }{2{\beta\cA}} \int \frac{d^{D+1}p}{(2\pi)^{D+1}} d^{D+1}x
	\,\tr \(
	G_W^{(0)} \ast \partial_{p_l} Q_W^{(0)} \ast
	G_W^{(0)} \ast \partial_{p_m} Q_W^{(0)} \ast
	G_W^{(0)} \ast \partial_{p_k} Q_W^{(0)}
	\).
	\label{cW}
\ee
Here $\beta = 1/T$ is the inverse temperature, $T$ is assumed to be small. Measure $dx$ contains both integration over spacial coordinates and over the imaginary time.
In the above expression we restored the $\ast$--product in the last factor using once again \Ref{no-star}. From now on we will omit the superscript $(0)$ for brevity.

Averaged conductivity (proportional to conductance) is given now by
\be
\bar\sigma_{mk} \equiv 	{\mathcal W}_{0mk}-	{\mathcal W}_{0km}
\label{bar_s}
\ee
It is anti-symmetrized with respect to the indices, and thus never gives rise to normal conductivity, only to the Hall one. This can be understood by noting that in the  conducting phase, where the longitudinal conductivity would be non-zero, the Green's function entering \Ref{cW} has a pole, thus giving rise to the uncertainty in the calculation of $\cW$. Therefore, the expression derived above is not applicable to this regime.

Depending on the dimensionality of the system \Ref{cW} can describe a nontrivial magnetotransport effects as well.
In the two dimensional case (i.e. with $2+1$D fermions embedded in $3+1$D space) for a system in the presence of electric field along the $ x_2$ axis we have
$$
\cJ_1^{(E)} = \rr{-}\frac{\cal N}{2\pi} E_2.
$$
Here
\be
{\cal N}
=  \frac{T \epsilon_{ijk}}{ {\cal A} \,3!\,4\pi^2}\,  \Tr
\[
{G}_W(x,p )\ast \frac{\partial {Q}_W(x,p )}{\partial p_i} \ast \frac{\partial  {G}_W(x,p )}{\partial p_j} \ast \frac{\partial  {Q}_W(x,p )}{\partial p_k}
\]_{A^{(E)}=0}
\label{calM2d230}
\ee
with $\Tr$ defined in \Ref{Tr}. This expression represents the averaged Hall conductivity.

In principle one may also consider the coefficient of proportionality between ${\mathcal J}$ and constant external magnetic field $B$ (now $\F^{(E)}_{ml}$ corresponds to magnetic field instead  of electric field). Notice that $B_{1,2}$ do not produce any current in $2+1$D via \Ref{cJ}. The only possible non-trivial component is for $B_3$, the magnetic field perpendicular to the plane,
\be
	{\mathcal J}_0^{(B)} = B_3
		\({\mathcal W}_{120}-	{\mathcal W}_{210}\)
		\sim\bar\sigma_{12} B_3,
\ee
i.e. in this case the magnetic field may induce the excess of electric charge density in the system (compared to the charge density in the absence of the external magnetic field).
In this expression we use the cyclic property of \Ref{cW}.
For the $3+1$D fermionic system the pattern is more complicated. Let us suppose that the magnetic field along $x_2$ axis is present, $B_2$, encoded by $F^{(E)}_{ml}$ of \Ref{cJ}. Then,
\be
	\cJ_1^{(B)} = B_2 \({\mathcal W}_{131}-	{\mathcal W}_{311}\) = 0
\ee
due to the cyclic property of  \Ref{cW}. However,
\be
	\cJ_2^{(B)} = B_2 \({\mathcal W}_{132}-	{\mathcal W}_{312}\)\label{B2W}
\ee
may, in principle, be nonzero. In practise, however, in the majority of equilibrium systems the total electric current without external electric field is zero (which is the content of the Bloch theorem \cite{ZZ2019_3}). In particular, in \cite{Zubkov2016a} it has been proved that the equilibrium chiral magnetic effect is absent, i.e. the corresponding coefficient in Eq. (\ref{B2W}) vanishes for the homogeneous system of lattice regularized Dirac fermions in the presence of chiral chemical potential.

$\cal N$ given by \Ref{calM2d230} or, equivalently, $\bar\sigma$ of \Ref{bar_s} is the topological invariant in phase space, as it can be readily checked in the way similar  to \Ref{deltaJ}. The variations under which it remains invariant must be ``small'', i.e.   should not change the behavior of the system at spatial infinity. It is important to observe that \Ref{calM2d230} is a more general invariant than the classical TKNN \cite{Thouless1982} being applicable to the non homogeneous and non-uniform systems.

\section{Other results}

\subsection{Connection to Kubo formula}

\label{SectKubo}
To restore the more familiar expression for the Hall conductivity, one may use Weyl representation in momentum space \Ref{A_W} substituted into \Ref{calM2d230}, and notice that the $p$-derivative acting on Weyl symbol becomes the sum of derivatives acting on the matrix elements of operator $\hat{\mathcal{G}}_k$ such that:
\be
\partial_{p_k} G_W = \Big(\hat{\mathcal{G}}_k\Big)_W
\ee
with
$$
\mathcal{G}_k(\cP,\cR)
	\equiv 	\(\partial_{\cP_k}+\partial_{\cR_k} \) G(\cP,\cR).
$$
The topological invariant becomes
\be
\cN =
	\frac{\epsilon_{abc}}{3! 4\pi^2\cA}
	\int (dl\, dk\, dp\, dq \\)^{D+1}
	\tr \[
	G(l,k) (\partial_{k_a}+\partial_{p_a}) Q(k,p)
	(\partial_{p_b}+\partial_{q_b}) G (p,q)
	(\partial_{q_c}+\partial_{l_c}) Q(q,l)
	\].
\label{NtoKubo}
\ee
%
For the non-interacting systems, Hamiltonian ${\cal H}$ has energy eigenstates $|n\rangle$: $\cH |n\rangle = \cE_n |n\rangle$. Then for $Q=\ii\omega - \cH$ we have
\be
Q(p,q) \equiv \braket {p| \hat{Q} |q }
= \( \delta^{(2)} (\bp-\bq) \ii \omega_p
- \braket{\bp| {\cal H} | \bq}\) 	
\delta(\omega_p-\omega_q),
\ee
where we restricted ourselves to the $2+1$D system, $p = (p_1,p_2,p_3) = (\bp ,\omega_p)$. At the same time
\be
G(p ,q) = \sum_{n} \frac{1}{\ii\omega_p-{\cal E}_n}
\braket{\bp| n} \braket{n | \bq} \delta(\omega_p-\omega_q).
\ee
Applying the two given above formulas in \Ref{NtoKubo}, the usual Kubo formula is restored,
\begin{eqnarray}
{\cal N}
	&=&  \frac{\ii\,(2\pi)^2}{8\pi^2\, {\cal A}}\,\sum_{n,k} \int_\dR \D\omega  \epsilon_{ij}\,
	\frac{\langle n| [{\cal H}, {\hat x }_i] | k \rangle  \langle k | [{\cal H}, {\hat x }_j] | n \rangle  }
	{(\ii\omega^{}-{\cal E}_n)^2 (\ii\omega^{}-{\cal E}_k)}
	\nonumber\\
&=&-\frac{2\ii\,(2\pi)^3}{8\pi^2\, {\cal A}}\,\sum_{n,k}   \, \epsilon_{ij}\,
	\frac{\theta(-{\cal E}_n)\theta({\cal E}_k)}{({\cal E}_k-{\cal E}_n)^2}
	\langle n| [{\cal H}, {\hat x }_i] | k \rangle
	\langle k | [{\cal H}, {\hat x }_j] | n \rangle,
\label{sigmaHH}
\end{eqnarray}
where the coordinate operator has the meaning of the derivative in $p$
$$
\hat{x}_j \Psi(\bp)
= \langle \bp|\hat{x}_j |\Psi\rangle
= i\partial_{p_j} \langle \bp|\Psi\rangle
= \ii \partial_{p_j} \Psi(\bp).
$$

\subsection{Introduction of interactions}

\label{SectInt}

According to \cite{ZW2019} Eq. (\ref{calM2d230}) gives the average Hall conductivity in the presence of the non-homogeneous magnetic field and non-homogeneous electric potential of impurities, but with the interactions neglected.
It is natural to suppose also, that Eq. (\ref{calM2d230}) remains valid in the presence of the electron-electron interactions.
One may consider following \cite{ZZ2019_2} the Euclidean lattice action in momentum  space
%
%
\be
S= \int  d^{D+1}p \bar{\psi}_{p}\hat{Q}(p,i\partial_p)\psi_{p} \rr{-}\alpha \int  d^{D+1}p d^Dq d^{D+1}k \bar{\psi}_{p+q}\psi_{p}\tilde{V}({\bf q})\bar{\psi}_{k}\psi_{q+k}.
\ee
For definiteness we may take the Coulomb interaction with $V({\bf x})=1/|{\bf x}|=1/\sqrt{x_1^2+x_2^2}$,
for ${\bf x}\not= {\bf 0}$. However, the consideration of the other types of interactions that occur due to the exchange by bosonic excitations is similar.
Then
$\tilde{V}({\bf q}) =\sum_{\bf x} \frac{e^{i{\bf q\cdot x}}}{\sqrt{x_1^2+x_2^2}}$.
%
The Coulomb interaction contributes to the self-energy of the fermions,
and the leading order contribution is proportional to $\alpha$.

The results of the calculations presented in \cite{ZZ2019_3} demonstrate, that the (averaged over the system area) Hall conductivity in the presence of inhomogeneous magnetic field, inhomogeneous electric field, and Coulomb interactions is proportional to the topological invariant in phase space of Eq. (\ref{calM2d230}). In the presence of interaction one simply has to substitute to Eq. (\ref{calM2d230}) the complete two-point Green's  function with the contribution of interactions included.  The present derivation of Eq. (\ref{calM2d230}) (see also \cite{ZW2019,FZ2019} where this derivation has been given in the absence of interactions) is valid for the gauge field potential that varies slowly at the distances of the order of lattice spacing. This corresponds to the values of magnetic field much smaller than thousands Tesla and the wavelengths much larger than several Angstroms.
 In the region of analyticity in $\alpha$ the Hall conductivity does not depend on $\alpha$ at all and is still given by the same expression as without Coulomb interactions.

\subsection{Hall conductivity in the presence of elastic deformations }
\label{SectElast}

In \cite{FZ2019} the technique discussed above was applied to the tight-binding model of graphene in the presence of both inhomogeneous magnetic field and nontrivial elastic deformations. The majority of the results obtained in \cite{FZ2019} may be applied to the family of two-dimensional honeycomb lattice materials (graphene, germanene, silicene, etc), and to the rectangular lattice crystals, see \cite{FZ2019}. In the mentioned cases the electrons may jump only between the nearest neighbors  and there is the $Z_2$ sublattice symmetry. The lattice consists of the two sublattices $\cO_1$ and $\cO_2$. For each $ \bx  \in \cO_1$ site $ \bx  + \bbj  \in \cO_2$ with fixed vectors $\bbj $, where $j = 1,2,...,M$. For the honeycomb lattice $M = 3$, for the $2D$ rectangular lattice $M = 4$, for the $3D$ rectangular lattice $M = 8$. 

Weyl symbol of lattice Dirac operator (i.e. the operator $\hat Q$ that enters the action $\sum_{\bx,\by } \bar{\Psi}_x Q_{\bx,\by } \Psi_y$) has been calculated in the presence of elastic deformations:
\be
{Q}_W =
\ii \omega-t  \sum_j
	\( 1- \beta u_{kl}({\bx})  \bj_k \bj_l \)
	\(\begin{array}{cc}
		0 & e^{\ii  (\bp  \bbj -{ A}^{(j)}({\bf r} ))} \\
		e^{-\ii (\bp  \bbj -{ A}^{(j)}({\bf r} ))} & 0
	\end{array}	\)\label{QWc}
\ee
where
 $u_{ij}$ is the tensor of elastic deformations while
 $$
	A^{(j)}(\bx) = \int_{\bx-\bbj /2}^{\bx + \bbj /2}  \bA(\by)d\by.
$$

\subsection{Solution of the Groenewold equation}
\label{SectGroenewold}

There are two possible ways of solving the Groenewold equation \Ref{GreoEqu}. First of all, the standard perturbation solution, build upon small perturbation of the Dirac operator
\be
	Q_W(x,p)\approx Q_W^{(0)}+Q_W^{(1)}+\ldots
\ee
Then, quite trivially,
\be
	G_w(x,p)\approx G_W^{(0)}- G_W^{(0)}*Q_W^{(1)}*G_W^{(0)}+\ldots,
\ee
where
\be
	G_W^{(0)}*Q_W^{(0)}=1.
\ee

Alternatively, a kind of gradient expansion can be constructed, if the modification of $Q_W$ cannot be written as a small perturbation. It is essentially based on the expansion of the pseudo-differential $*$-operator \Ref{star} in powers of
$$
\overleftrightarrow{\Delta} = \frac{\ii}{2}
	\( \overleftarrow{\partial}_{x}\overrightarrow{\partial_p}-\overleftarrow{\partial_p}\overrightarrow{\partial}_{x}
	\).
$$
Constructing the appropriate iterative scheme and performing resummation of the latter, see \cite{SZ2018}, we obtain the following expression for Wigner transformation of electron propagator:
{\begin{equation}\begin{aligned}
{G}_W(x,p )
=&\sum_{k=0}^\infty  \, \underbrace{\Big[...\Big[Q^{-1}_W(1- e^{\overleftrightarrow{\Delta}}) Q_W\Big] Q^{-1}_W (1-e^{\overleftrightarrow{\Delta}}) Q_W\Big]... (1-e^{\overleftrightarrow{\Delta}}) Q_W \Big]} Q^{-1}_W\\
& \hspace{5cm} \emph{k\rm\ brackets}
\label{GWc}.
\end{aligned}\end{equation}}
It is also valid in the presence of slowly varying magnetic field and arbitrary elastic deformations \cite{FZ2019,Fialkovsky2020b}.

Correspondingly, in \cite{FZ2019} it has been shown that the Hall conductivity in these  systems is given by the same expression of Eq. (\ref{calM2d230}) with the above presented expressions for $Q_W$ and $G_W$.

\section{Conclusions}

\label{SectConcl}

To conclude, in this paper we review the results obtained by the group of the authors and published earlier in the series of papers \cite{SZ2018,ZW2019,ZZ2019_2,FZ2019}. Below we summarize the main obtained results

\begin{enumerate}

\item{}
Wigner-Weyl calculus for the lattice models has been developed for the case of the slowly varying external fields \cite{SZ2018,FZ2019}. Wigner transformation of the two-point Green's  function and Weyl symbol of lattice Dirac operator are defined. The principal way to calculate both of them is proposed.

\item{}

It is shown that the Hall conductivity averaged over the system area in the $2+1$ D systems is proportional to the topological invariant of Eq. (\ref{calM2d230}) \cite{ZW2019} for the case of varying magnetic field and varying electric potential of impurities.

\item{}

The influence of interactions on the Hall conductivity is investigated. It has been shown, that it is still given by Eq. (\ref{calM2d230}) with the complete two-point Green's  function substituted instead of the non-interacting one \cite{ZZ2019_2}.

\item{}

The influence of elastic deformations on the Hall conductivity in graphene-like materials has been investigated. It was shown, that it is still given by Eq. (\ref{calM2d230}) \cite{FZ2019,Fialkovsky2020b}. The corresponding expression for the Weyl symbol of lattice Dirac operator has been calculated. The iterative solution of the Groenewold equation for the Wigner transformation of the Green's  function has been given \cite{SZ2018,FZ2019}.

\end{enumerate}

It is worth mentioning again, that the original TKNN invariant has been derived for the uniform magnetic field (constant both as a function of time and space coordinates). The expression for the Hall conductivity proposed in the mentioned series of papers is an extension of the TKNN invariant to the case of varying (in space) magnetic fields, or otherwise inhomogeneous systems. Therefore, its consideration is important. The non–renormalization of the Hall conductivity (given by the original TKNN invariant) by interactions has been discussed earlier. But it was limited by the case of constant magnetic fields {as well}. We gave the proof that the QHE conductivity (given by our extension of the TKNN invariant) is robust to the introduction of interactions in the case of varying magnetic field. This result has never been obtained in the past, to the best of our knowledge as well as the result on the influence of elastic deformations on the Hall conductivity in the presence of varying magnetic field.

The mathematical form of the topological invariant in phase space discussed here is somehow similar to the one of the topological invariant in momentum space composed of the two–point Green's  function. The latter topological invariant and its variations are used widely (see \cite{Volovik2003a}). Now the Green's  function is substituted by its Wigner transformation depending on both space coordinates and momentum. The ordinary products are therefore changed to the Moyal (star) product, thus leading to the beautiful mathematical structure. The Green's  functions with larger number of legs do not contribute to the Hall conductivity.

M.A.Z. is indebted for valuable discussions to G.E.Volovik.





	


\bibliography{wigner3,cross-ref}

\end{document}